# Protection of Guizhou Miao Batik Culture Based on Knowledge Graph and Deep Learning

**Huafeng Quan[1], Yiting Li [1,*], Dashuai Liu[2], Yue Zhou[3]**

[1] College of Big Data and Statistics, Guizhou University of Finance and Economics,Guiyang 550025, China
[2] School of Art Design and Media, East China University of Science and Technology, Shanghai 200237, China
[3] Guizhou Electronic Technology College，Guiyang, 550025, China
***** Correspondence: yitingli.cn@mail.gufe.edu.cn (Y.L.)

Abstract: In the globalization trend, China's cultural heritage is in danger of gradually disappearing. The protection and inheritance of these precious cultural resources has become a critical task. This paper focuses on the Miao batik culture in Guizhou Province, China, and explores the application of knowledge graphs, natural language processing, and deep learning techniques in the promotion and protection of batik culture. We propose a dual-channel mechanism that integrates semantic and visual information, aiming to connect batik pattern features with cultural connotations. First, we use natural language processing techniques to automatically extract batik-related entities and relationships from the literature, and construct and visualize a structured batik pattern knowledge graph. Based on this knowledge graph, users can textually search and understand the images, meanings, taboos, and other cultural information of specific patterns. Second, for the batik pattern classification, we propose an improved ResNet34 model. By embedding average pooling and convolutional operations into the residual blocks and introducing long-range residual connections, the classification performance is enhanced. By inputting pattern images into this model, their subjects can be accurately identified, and then the underlying cultural connotations can be understood. Experimental results show that our model outperforms other mainstream models in evaluation metrics such as accuracy, precision, recall, and F1-score, achieving 99.0%, 99.0%, 98.9%, and 99.0%, respectively. This research provides new ideas for the digital protection of batik culture and demonstrates the great potential of artificial intelligence technology in cultural heritage protection.

---

## 1. Introduction

Batik is an ancient manual dyeing technique found in many countries around the world, such as China, Indonesia, Malaysia, Singapore, India, and Japan. Influenced by multiple factors like social culture, geographical environment, and economy, batik from different regions presents distinct characteristics and styles in patterns, colors, and compositions [1-3]. Batik in China has a long history, dating back to the Qin and Han dynasties over two thousand years ago. It gained popularity during the Six Dynasties period and reached its peak in the Sui and Tang dynasties. However, with the passage of time and social development, this ancient handicraft has gradually declined. The impact of modernization, coupled with a lack of inheritors, has caused Chinese batik to face technological and cultural loss. Its development level lags far behind countries like India, Indonesia, and Malaysia, and it is on the verge of disappearing. This not only means the end of a traditional craft, but also the loss of related historical, cultural, social, and economic information.

In-depth research on the connotation of batik patterns is of great significance for understanding their historical evolution and cultural connotation, and can provide valuable references and inspiration for related fields such as art design and anthropology. At the same time, exploring new ideas and methods of digital preservation through the introduction of advanced computer technologies such as knowledge graphs, natural language processing (NLP), and deep learning can provide a systematic digital storage space for batik culture. Furthermore, constructing computational models can enable the automatic classification and identification of batik patterns, thereby revealing their intrinsic laws and relationships, and providing new opportunities for the inheritance and innovation of batik. This is of great significance for sustaining the vitality of batik culture and promoting the digital protection of cultural heritage.

Benefiting from the advantages of isolated terrain, Guizhou Miao batik has been relatively well protected and developed compared to other regions in China, enjoying the reputation of "the hometown of batik" [4-6]. The Miao people, an ethnic group with only spoken language but no written script, have made batik patterns an important carrier for inheriting history, religion, beliefs, customs, etc [7]. With its mysterious style, beautiful patterns, clear themes, heavy connotations, and rich subjects, Guizhou Miao batik patterns have thrived through generations and become one of the most representative Chinese cultural heritages.

The production principle of batik is to use the hydrophobicity of wax to prevent certain areas of the fabric from being dyed, thus creating patterns. The materials used in the production of batik mainly include wax (e.g. beeswax, paraffin, etc.), white cotton cloth, indigo, and copper knives. As shown in Fig. 1, the production of batik mainly involves four steps. (i) Melting wax. The solid wax is heated to the melting point, which turns it into a liquid state and provides plasticity for the subsequent drawing process. (ii) Wax painting. A copper knife is used to scoop the melted wax and draw various patterns on the white cloth. (iii) Dyeing. The wax-painted cloth is immersed in indigo water, allowing the dye to penetrate the areas not protected by wax. (iv) Removing wax. The dyed cloth is placed in boiling water, where the wax melts at high temperatures and is washed away by the water flow, resulting in blue-and-white batik patterns. Indigo is fermented from the leaves of bluegrass, a common plant in the mountainous regions of Guizhou. It is worth mentioning that bluegrass is not only a dyestuff but also a medicinal plant. Its properties of resisting humid climates and preventing mold growth help protect the texture and color stability of batik fabrics under the humid climatic of southern China.

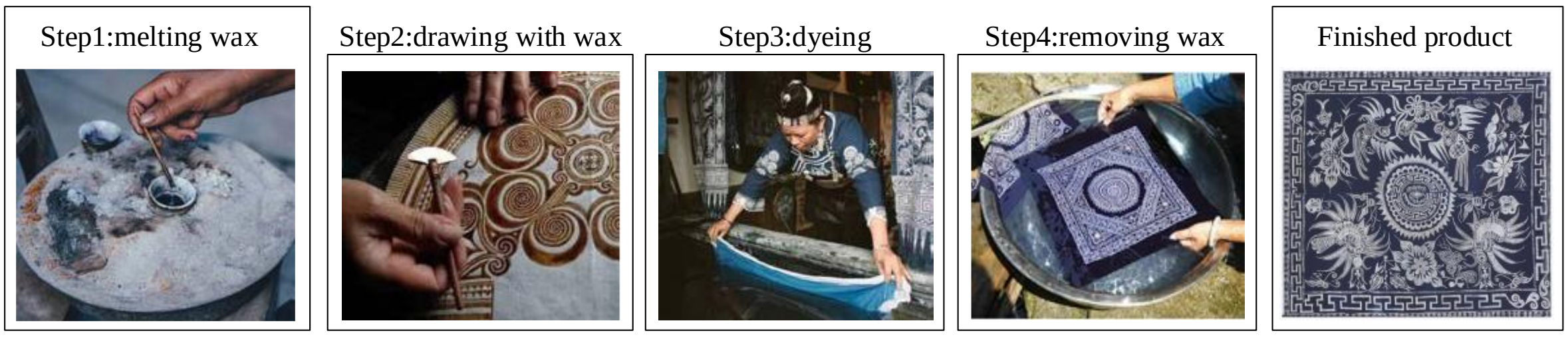

Fig. 1 The production process of batik

Batik fabrics are mainly used for decorating daily life, expressing beliefs, and conveying emotions. They are commonly used as bedsheets, door curtains, window curtains, skirts, and accessories. The patterns on the fabric are varied and have special meanings, some of which are illustrated in Fig. 2. The subjects of the patterns include plants, animals, and geometric shapes. Different patterns have different meanings. For example, the butterfly pattern contains the Miao people's mythical story of the "蝴蝶妈妈" (Butterfly Mom), the fish pattern means praying for having children, and the fish-bird pattern represents the marital harmony. In addition to the meaning of goodness, Miao culture also has certain taboos. To avoid conflicts between different cultures, the use of taboo patterns needs to be treated with caution. For example, the combination of the curled grass pattern and the horseshoe pattern represents guiding the way for the deceased.

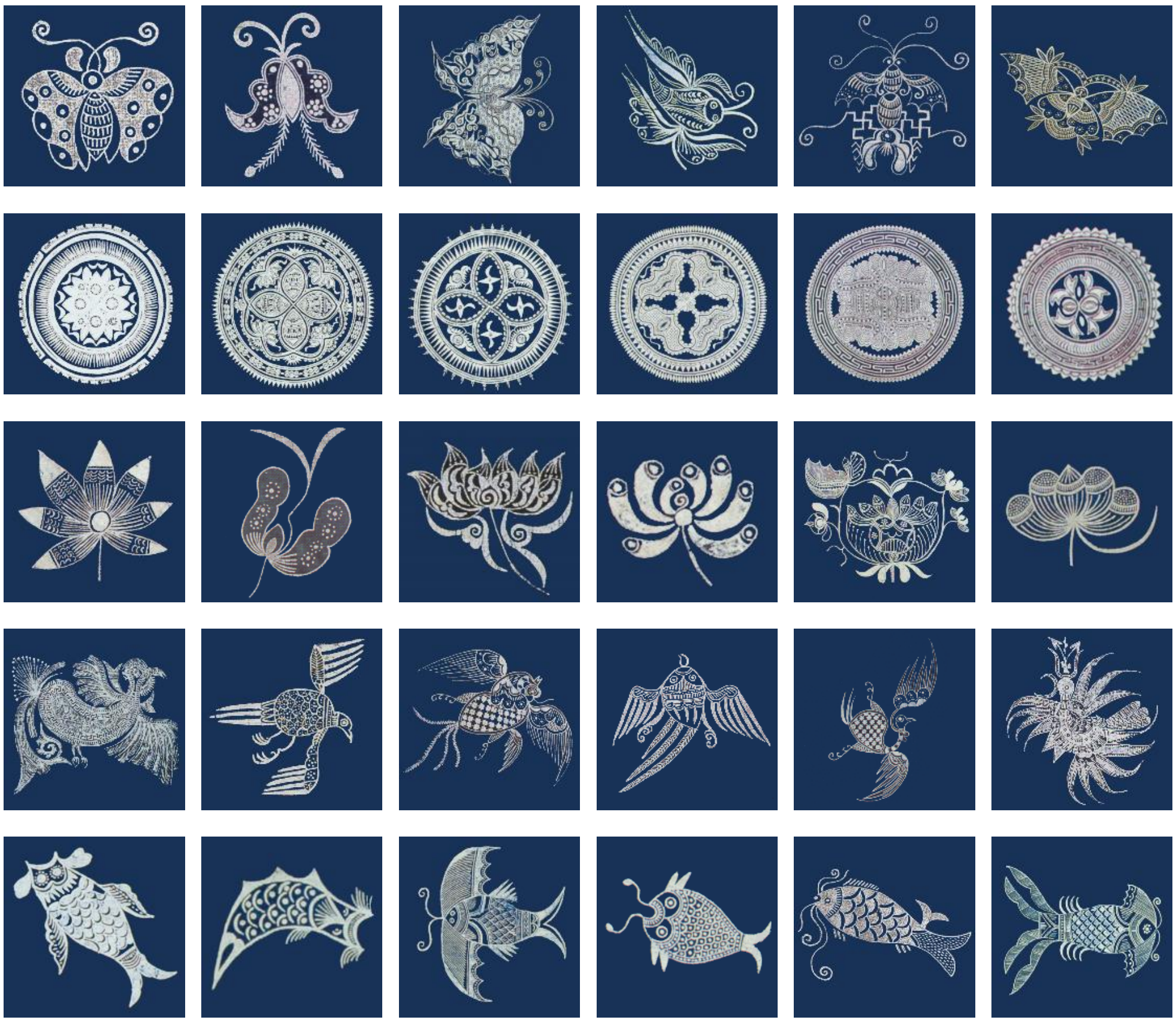

Fig. 2. Batik patterns.

Batik has attracted widespread attention due to its high aesthetic and cultural value. In the innovative design of batik patterns, TIAN et al. [8] proposed an automatic generation method for batik patterns based on fractal geometry. By introducing fractal theory, they realized the automatic generation and transformation of batik patterns, providing new ideas for batik pattern design. LV et al. [9] introduced an interactive genetic algorithm (IGA-BPFIF) into the innovative design of batik patterns. By using user interaction feedback to guide the optimization process of the genetic algorithm, they generated new batik patterns that conform to users' aesthetic preferences. In their later research [10], they also proposed an improved collaborative filtering algorithm to recommend batik patterns. By analyzing user preferences, the algorithm recommends batik patterns of interest to users. Hu et al. [11] proposed a generative design method for batik patterns based on shape grammars. Considering the low abstraction level of traditional shape grammars in batik design, DING et al. [12] improved it based on comprehensive predicate grammar encoding and used particle swarm optimization to optimize the parameters of predicate shape grammars, enhancing the diversity and artistry of batik pattern generation.

A deep understanding of the cultural background and spiritual connotation of batik patterns is the foundation for protecting them [13]. In the exploration of Miao batik culture, researchers have focused on studying the migration history, religious beliefs, aesthetic characteristics, and the meanings of specific patterns. For example, Zhennan and Yahaya [7] introduced in detail the origins and meanings of typical patterns such as fish, birds, butterflies, dragons, and plants, and summarized the Miao people's aesthetic preference for symmetry and fullness. In addition, some scholars have also discussed the artistic changes and regional differences of batik patterns [14]. Although these studies are highly practical, they have not organized these isolated knowledge into a systematic framework, resulting it difficult to accumulate, disseminate, and reuse. Considering the complexity, diversity, and sensitivity of culture, in this paper, we construct a knowledge graph to effectively store, organize, and manage batik pattern knowledge.

In terms of feature extraction from batik images, chen and cheng [15] proposed an extraction method based on morphological processing techniques and the canny algorithm, obtaining independently editable pattern contours. In the retrieval of batik patterns, YUAN et al. [16] proposed a method combining global and local features. First, global features and local features are extracted through Zernike moments (ZMs) and curve transformations, respectively. Then, these two types of features are combined by matching weighted bipartite graphs, and the visual similarity between patterns is calculated through supervised distance measurement. Finally, similar shape patterns are retrieved. Since Miao batik images are highly abstract, it is difficult to accurately distinguish their subjects through simple understanding. Moreover, culture is highly sensitive, and once incorrectly identified and used, it may cause very serious consequences. Therefore, in this paper, we employ an improved ResNet34 model to extract features from batik images and automatically classify them.

This paper proposes to establish a dual-channel mechanism between batik images and culture to reduce the difficulty of understanding and applying batik knowledge. In the direction from "culture to image," users can retrieve pattern images by the names, meanings, and taboos of the patterns. For example, when designers want to use patterns that represent "having many children" in their products, they can search for this keyword in the knowledge graph. The system will filter out patterns that match the meaning, such as pomegranate patterns and fish patterns, and display a large number of vivid image examples. The rich batik image resources not only stimulate designers' creative inspiration but also endow the products with cultural connotations. In the direction from "image to culture," pattern images are input into the classification model to identify the subject of the pattern and further explore its underlying cultural connotations. For example, when designers come across a beautiful batik pattern and want to use it in their products, but do not know the pattern's name, cultural meaning, historical origin, usage taboos, and other information, they dare not use it rashly. At this time, they can upload the pattern image to the model, which will identify the subject of the pattern, such as butterfly pattern or dragon pattern, and then explore its culture through the knowledge graph. This process of starting from images, exploring, and perceiving the underlying cultural connotations can bring ordinary people closer to traditional handicrafts and enhance their cognition of cultural heritage, strengthening their willingness to apply batik elements in their work and life. The main contributions of this paper can be summarized as follows:

(i) For the organization and management of batik pattern knowledge, we adopt a method that integrates NLP and knowledge graphs to construct a Batik Pattern Knowledge Graph (BPKG). Entities and relationships of batik patterns are automatically extracted from a large amount of textual data and semantically associated. Compared with traditional knowledge base construction methods, our approach can organize and manage domain knowledge more comprehensively and systematically.

(ii) For batik pattern classification, we propose an improved ResNet34 model. By embedding average pooling and convolutional operations into the residual blocks and introducing long-range residual connections, this structure can enhance the network's feature extraction and representation ability and effectively alleviate the gradient vanishing problem. Compared with traditional convolutional neural networks (CNN) and original residual networks, our model achieves superior performance on the batik pattern classification task.

(3) We created a large-scale batik pattern image dataset containing 12,249 images. The dataset covers five subjects. This is the largest and most complete image dataset in the field of Chinese batik, which is of great significance in promoting the digitalization and intelligent development of batik culture.

In summary, this paper conducts research on the construction of BPKG, the design of classification model, and the creation of batik dataset, providing new ideas and methods for the digital protection and inheritance of batik culture, with strong theoretical and practical value. Advanced technologies often take longer time to become popular in the cultural heritage field [17], the innovation of this paper is here, that is, to introduce these advanced computer technologies into batik, exploring the integration of traditional culture and artificial intelligence (AI) technology. The research results can not only promote the digital development of batik culture but also provide new perspectives and approaches for the intelligent protection and inheritance of other traditional culture heritages, which has exemplary significance and promotional value.

The overall structure of this paper is as follows. In Section 2, we review the related works. In Section 3, we introduce the knowledge graph and ResNet model in detail. In Section 4, we present the specific research work, first constructing the batik pattern knowledge graph, then improving the ResNet34 model, and finally applying it to batik image classification. Section 5 provides a summary.

## 2. Related works

### 2.1. Knowledge Graph

The concept of knowledge graph was first proposed by Google in 2012. It is a method of graphically representing knowledge and its relationships. Formalized knowledge representation methods include predicate logic, frame-based, ontology-based, etc. Among them, the predicate logic method uses predicates and logical connectives to describe concepts and relationships; the frame-based method uses frame structures to organize and describe concepts; and the ontology-based method uses elements such as classes, instances, attributes and relationships to describe domain-specific knowledge. In comparison, the ontology-based method can ensure the uniqueness of knowledge understanding in the process of transfer and sharing, and can meet the requirements of diverse knowledge types and complex semantic relations, making it a widely used knowledge representation method.

In the automatic construction of knowledge graphs, with the help of NLP and machine learning techniques, some studies have realized the semi-automatic extraction and fusion of knowledge, greatly reducing the time and cost required for semantic processing and graph construction. The key technologies involved include entity extraction, relation extraction, and knowledge fusion. Entity extraction involves extracting entities and their attributes from text, commonly used methods include rule-based and machine learning-based methods [18-19]. Rule-based methods identify entities by defining a series of matching patterns and rules, such as regular expressions, dictionary matching, etc. Machine learning-based methods treat entity extraction as a sequence labeling problem and automatically identify entities by training sequence labeling models (e.g., HMM, CRF, BiLSTM-CRF, etc.).

Relation extraction involves identifying the semantic relationships between entities from text, such as "located in" and "belongs to". Commonly used methods include pattern matching, key word extraction-based, and machine learning [20-22]. The pattern matching method extract relationships by defining relation templates and trigger words. The key word extraction method determine the relationships between entities by identifying the key verbs in sentences. The machine learning method treat relation extraction as a classification problem and predict the relation types of entity pairs by training classifiers (e.g., CNN, RNN, attention mechanisms, etc.).

Knowledge fusion is the integration of entities and their relationships from different data sources into a unified knowledge graph to create a more comprehensive and richer knowledge base. It mainly includes two tasks: entity linking and knowledge merging [23-24]. Entity linking is to link entity mentions in text to existing entities or create new entities. Commonly used methods include collaborative filtering, random walks, rule-based and deep learning methods. Knowledge merging is to identify equivalent entities, relationships and attributes in different data sources and merge them. Commonly used methods include similarity-based clustering, logical rule-based reasoning, ontology matching-based techniques, etc.

According to different functions and application scenarios, knowledge graphs can be divided into two categories: general knowledge graphs and domain knowledge graphs. General knowledge graphs aim to cover a wide range of fields and topics, while domain knowledge graphs focus on specific fields, industries, or topics, with higher requirements for the depth and quality of knowledge.

In the general domain, commonly used knowledge graphs include Wikidata, DBpedia, Freebase, YAGO, and Google Knowledge Graph [25-26]. Advances have also been made in domain knowledge graphs, such as medicine, social interaction, agriculture, aviation, product design, cultural heritage, progress has also been made [27-29]. For example, due to the large scale and fragmented nature of knowledge in the cultural heritage domain, which is not conducive to knowledge dissemination and management, Carriero et al. [30] established the Italian cultural heritage KG ArCo. Similarly, Dou [31] et al. used the Bi-GRU model to extract entity relations and constructed a knowledge graph of China's Twenty-Four Solar Terms. Fan [32-34] et al. have conducted a series of studies on China's intangible cultural heritage. They constructed unimodal and multi-modal knowledge graphs, integrating multi-source heterogeneous data such as text, images, and audio, providing new ideas for the multi-level and multi-angle presentation of intangible cultural heritage knowledge. Yang [35] et al. developed a public cultural knowledge graph platform that supports various application functions such as knowledge query, reasoning, clustering, ranking, similarity, classification, and visualization.

Although knowledge graphs have been widely applied in the cultural heritage domain, research on batik culture is still very limited. As one of the representative projects of China's intangible cultural heritage, batik culture contains rich knowledge in history, art, folklore, and other aspects. However, most of this knowledge exists in unstructured forms such as text and images, which is not conducive to effective knowledge management and utilization. Therefore, constructing a batik knowledge graph to structurally represent and organize batik-related knowledge is of great significance for the digital protection, dissemination, and innovative development of batik culture. Based on the batik knowledge graph, intelligent batik craft display systems, intelligent tools to assist batik pattern design, and batik culture popularization platforms for the public can be developed to effectively promote the dissemination and application of batik.

## 2.2. Image Classification

Image classification is an important task in computer vision. Its goal is to assign input images to predefined categories or labels. Early image classification methods mainly relied on handcrafted features and traditional machine learning algorithms. These methods usually include two parts: feature extraction and classification. In the feature extraction stage, commonly used methods include grayscale histograms, edge histograms, texture features (such as Gabor filters), color histograms, etc. The extracted features are usually low-level or mid-level. In the classification stage, the extracted features are input into traditional statistical methods or machine learning algorithms, such as support vector machines (SUM), k-nearest neighbor (KNN) algorithms, decision trees, random forests, etc. [36-37], to train classifiers.

However, traditional methods have some limitations, such as the dependence of feature engineering on expert knowledge and the limited feature representation capability. With the rise of deep learning, methods based on deep neural networks have gradually replaced traditional methods and become the mainstream technology in this field [38]. Unlike traditional methods, deep learning methods can automatically learn hierarchical feature representations without the need for handcrafted features. In these methods, feature extraction and classification tasks are usually performed end-to-end. End-to-end means that the entire process from input images to output classification results is completed in the same neural network, without the need for separate feature extraction and classification steps. Common deep learning image classification models include CNN, ResNet, RNN, and long short-term memory networks (LSTM). In addition, some studies attempt to combine traditional methods with deep learning methods, using deep networks to extract features and then using traditional machine learning algorithms for classification [39], to leverage the advantages of both types of methods.

In recent years, image classification techniques have shown broad application prospects in the field of cultural heritage [40-41], especially in the classification of patterns and style recognition. Ding et al. [42] proposed a nearest-neighbor method-based image classification model for She clothing, which integrates the texture features and spatial layout features of the clothing texture to improve the classification accuracy. In their follow-up research [43], they introduced CNN and designed a color feature fusion strategy optimized by the flower pollination algorithm, making the classification model simultaneously consider multiple visual features such as color, texture, and space of clothing, achieving better classification results. Kong et al. [44] focused on the pattern classification problem of Yao ethnic clothing and brocade, proposing a multi-target classification method based on Faster R-CNN. This method can not only simultaneously identify multiple patterns in an image but also synchronize pattern classification and localization, providing support for fine-grained analysis and application of patterns. In addition, they also developed a corresponding mobile application, allowing users to understand the meaning of patterns through scanning, greatly promoting the dissemination of Yao culture. To improve the classification effect of CNN on specific datasets, Jia and Liu [45] improved CifarNet and proposed the CalicoNet model, which classified 12 types of blue calico patterns. Furthermore, Fang et al. [46] combined images with text descriptions and proposed a multi-modal image classification model MICMLF, which was validated on the New Year Print and Clay Figurine datasets.

In the research of batik image or pattern classification, scholars have also carried out some work. Nurhaida et al. [1] made an early attempt to apply machine learning to Indonesian batik image classification. They proposed a method based on the gray-level co-occurrence matrix and KNN classifier, achieving an accuracy of 85% on a self-built dataset. Danis et al. [2] designed a classification method based on CNN for Japanese batik images, obtaining an accuracy of 90.14% on a self-built dataset. Pramerdorfer et al. [3] adopted the idea of transfer learning, using the pre-trained VGG16 model to classify Indonesian batik images, achieving an accuracy of 89% on a public dataset. These studies show that both traditional machine learning methods and deep learning-based methods have made positive progress in the batik image classification task. However, reviewing the existing literature, we find that research on Chinese batik in the image classification field is still relatively scarce, with a large gap compared to other countries. In our survey, no research work specifically targeting Chinese batik image classification has been found. One reason for this situation may be that the digitization level of Chinese batik is relatively low, and large-scale, high-quality image datasets have not been constructed, which to a certain extent limits the application of machine learning and deep learning-based classification methods. This situation reflects, to a certain extent, that the development level of Chinese batik culture still lags behind other countries, and there is an urgent need to strengthen scientific and technological innovation to promote the digital protection and intelligent inheritance of Chinese batik culture.

## 3. Methods

### 3.1. Research Framework

To help understand and apply batik knowledge, we comprehensively utilize knowledge graphs, NLP techniques, and an improved ResNet34 model to establish a dual-channel mechanism batik images and culture. This mechanism integrates the image features and cultural semantic information of batik patterns. On the one hand, it can retrieve relevant images based on cultural information; on the other hand, it can automatically identify the category of a batik image and then associate it with the corresponding cultural semantics. Our research framework is shown in Fig. 3. In the knowledge graph part, we first take the text data from the literature as the object, then use NLP techniques such as Word2vec and dependency parsing to automatically extract the key entities (such as "butterfly pattern" and "Butterfly Mom") and their relations (such as "mean" and " belong to") in the batik domain. Next, we refer to the seven-step method to construct the ontology, perform semantic modeling and normalization processing on the extracted entities and relations, forming a structured and semantic batik knowledge graph. Finally, we use the Neo4j graph database to store the knowledge and provide visualization and query interfaces. In the batik image classification model part, we first construct a large-scale batik pattern image dataset through data collection and annotation. Then, based on the classic ResNet34 model, we propose an improved structure that enhances the model's ability to extract and represent the features of batik patterns. Finally, we train and evaluate the improved ResNet34 model on the constructed batik pattern image dataset and apply it to the batik image classification task.

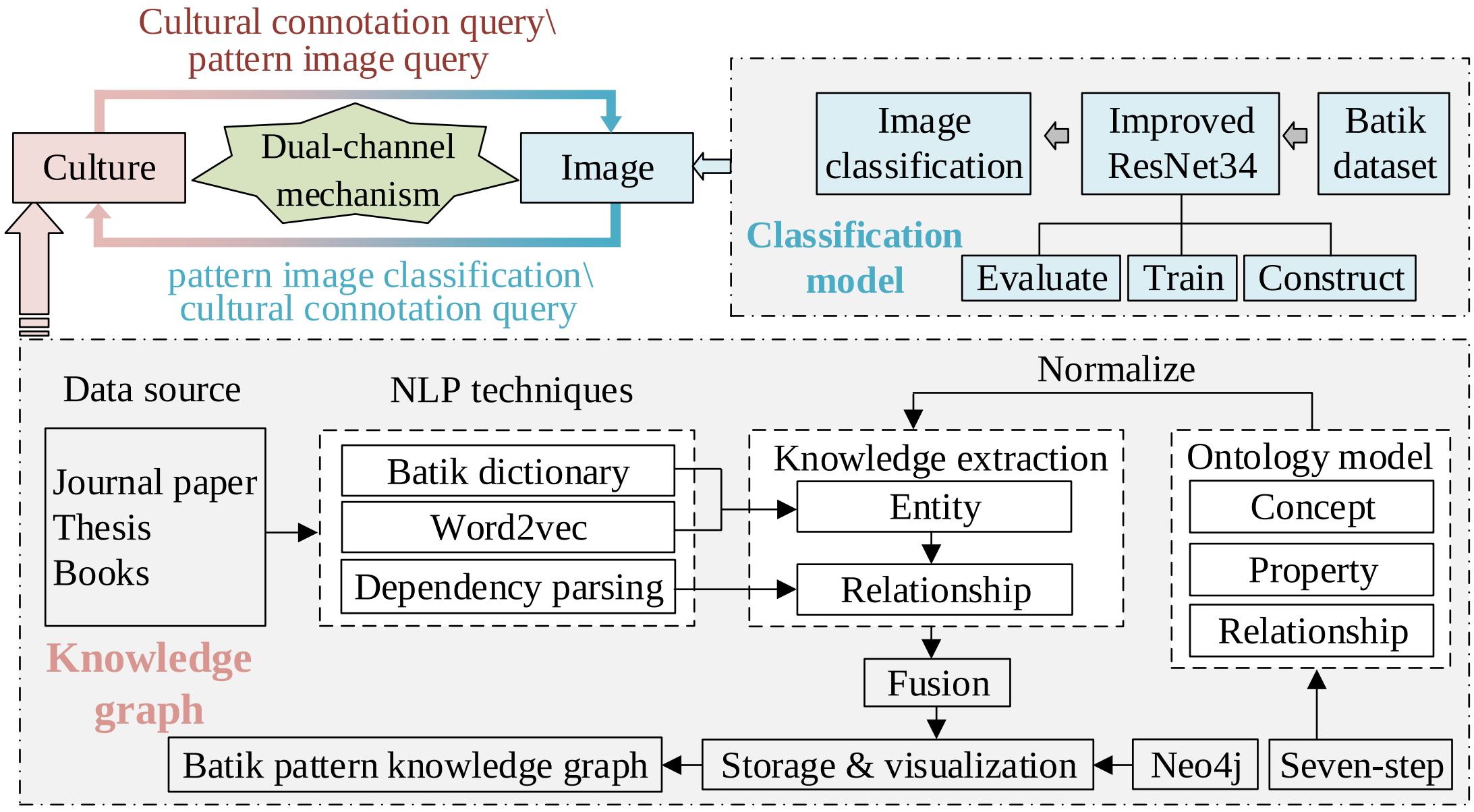

Fig. 3. Research framework.

### 3.2. Word Embedding

Word embedding is a technique that maps words to vectors, representing the semantic relationships between them through their positions in the vector space. Compared to traditional one-hot representation, word embeddings can better capture the semantic information of words and have been widely used in NLP tasks. This paper uses the Word2Vec model to train word embeddings, which includes two architectures: Continuous Bag-of-Words (CBOW) and Skip-gram models. Fig. 4 shows the structures of these two models, which consist of an input layer, a projection layer, and an output layer.

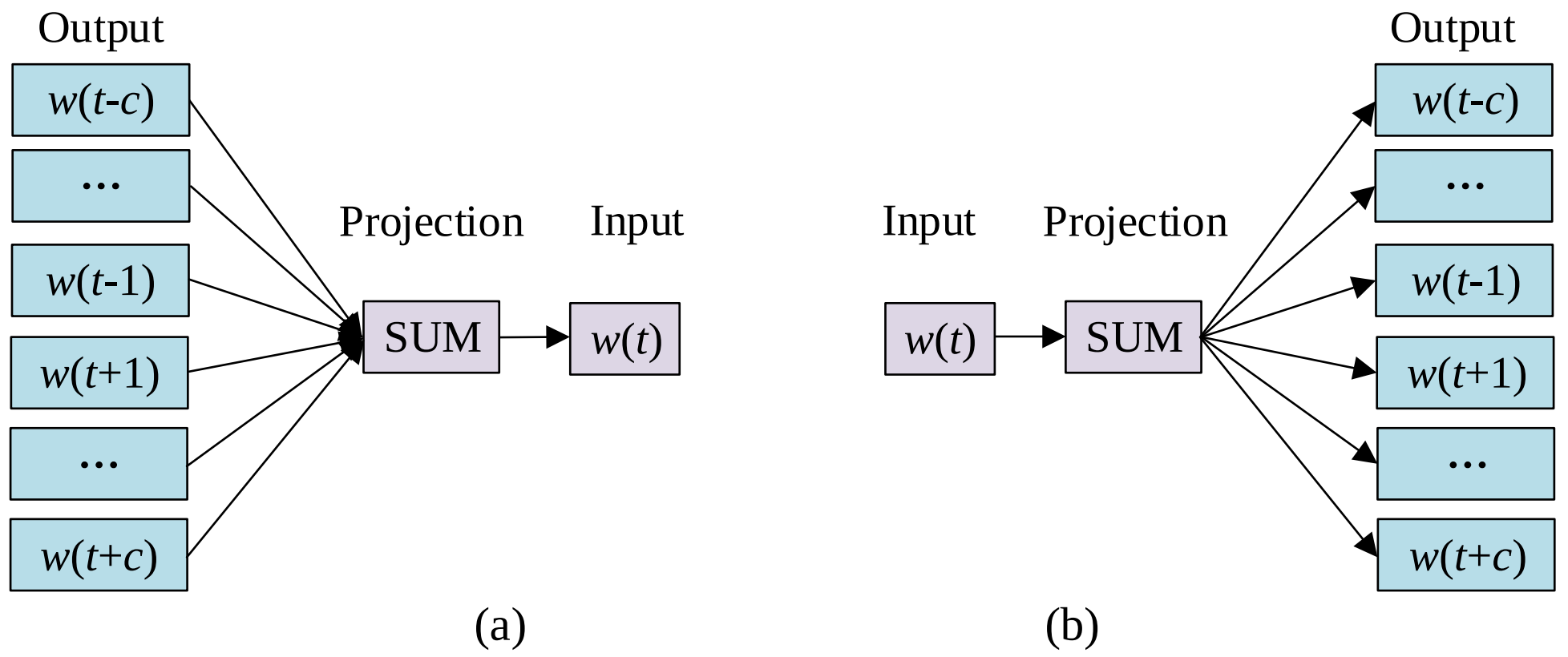

Fig. 4. Two training models of Word2Vec. (a) CBOW model (b) Skip-gram model

The training goal of CBOW is to predict the center word through the context words. Specifically, given a sequence of words $w_{t-c},\cdots,w_{t-1},w_{t+1},\cdots,w_{t+c}$ of length 2*c*+1. Where $w_t$ is the center word, *c* is the context window size, and *T* is the total number of words in the corpus, the objective function of CBOW can be expressed as:

$$L_{CBOW}=\frac{1}{T}\sum_{t=1}^{T}\log p(w_t\mid w_{t-c},\cdots,w_{t-1},w_{t+1},\cdots,w_{t+c}) \quad (1)$$

where $p(w_t\mid w_{t-c},\cdots,w_{t-1},w_{t+1},\cdots,w_{t+c})$ represents the conditional probability of the center word $w_t$ given the context words $w_{t-c},\cdots,w_{t-1},w_{t+1},\cdots,w_{t+c}$.

Skip-gram predicts the context words through the center word. Its objective function is:

$$L_{SK}=\frac{1}{T}\sum_{t=1}^{T}\sum_{-c\le j\le c,j\ne 0}\log p(w_{t+j}\mid w_t) \quad (2)$$

where $p(w_{t+j}\mid w_t)$ represents the conditional probability of the context word $w_{t+j}$ given the center word $w_t$.

During training, CBOW and Skip-gram learn the vector representations of words by maximizing the objective function. However, since the size of the corpus is usually large, directly computing the conditional probability is costly. To improve training efficiency, the Word2Vec model employs a strategy called Negative Sampling to approximate the conditional probability.

The idea of Negative Sampling is that for each positive sample $(w_t,w_c)$, *K* negative samples $(w_k,w_k)\mid k=1,\cdots,K$ are randomly sampled, and then the following function is maximized:

$$\text{loss}=\log\sigma(v_{w_c}^{'T}v_{w_t})+\sum_{k=1}^{K}\mathrm{E}_{w_k\sim P_n(w)}[\log\sigma(-v_{w_k}^{'T}v_{w_t})] \quad (3)$$

where $w_t$ represents the center word, $w_c$ represents the context word, $v_{w_t}$ and $v_{w_c}^{'}$ represent the word embeddings of the center word and context word respectively; $v_{w_c}^{'T}$ represents the transpose of $v_{w_c}^{'}$, which is used for vector dot product computation to calculate the similarity between $v_{w_t}$ and $v_{w_c}^{'}$; $P_n(w)$ is the negative sampling distribution, usually chosen as the 3/4 power of the word frequency distribution in the training corpus; $\mathrm{E}_{w_k}\sim P_n(w)[\cdot]$ represents the expected value of computing the expression inside the bracket $[\cdot]$ for the negative sample word $w_k$ sampled from the distribution $P_n(w)$; $\sigma(\cdot)$ is the sigmoid function, which maps a real value to the interval (0,1) and can be interpreted as a probability value.

3.3. Knowledge Graph

A knowledge graph is a structured knowledge representation that describes the relationships between entities in the form of a graph. In a knowledge graph, knowledge is stored and organized as triples of "entity-relation-entity", represented as: $G = (E, R, S)$. Here, $E$ is the set of entities, containing $n$ different entities, i.e., $E=\{e_1, e_2,\ldots,e_n\}$. Each entity represents a concrete object or abstract concept, such as "animal pattern" and "butterfly pattern". $R$ is the set of relations, containing $m$ different relations, i.e., $R=\{r_1, r_2,\ldots,r_m\}$. Each relation represents the connection between two entities, such as "belong to" and "symbolize". $S \subseteqq E*R*E$ is the set of triples. Each triple $(e_i, r_k, e_j)$ represents a piece of knowledge, indicating that there exists a relation $r_k$ between entity $e_i$ and entity $e_j$. For example, the triple (butterfly pattern, symbolize, respecting ancestors) represents "butterfly pattern symbolize respecting ancestors".

Knowledge graphs are usually stored using graph databases, which is a NoSQL database that uses a graph structure (consisting of nodes and edges) to store and query data. Entities are stored as nodes of the graph and relations are stored as directed edges connecting two nodes. Each node and edge can have their own properties to describe additional information about the entities and relations. Graph databases provide support for the storage, querying, and visualization of knowledge graphs.

### 3.4. Improved ResNet34 Model

In deep learning, CNN has become the mainstream method for image classification tasks. By stacking multiple convolutional layers and pooling layers, it can automatically learn and extract hierarchical features from images and integrate the feature representation with the classifier in an end-to-end architecture. However, traditional CNN models often encounter the problem of gradient vanishing or exploding when the network deepens to a certain extent, making it difficult to improve the classification performance.

To solve this problem, He et al. [47] proposed the residual network (ResNet). The idea of ResNet is to introduce the concept of identity mapping, i.e., adding "shortcut connections" that directly pass input information to the output in the network. In this way, even when the network is very deep, gradients can be directly propagated to shallower layers, thus alleviating the problem of gradient vanishing. The design principle of the ResNet is shown in Fig.5.

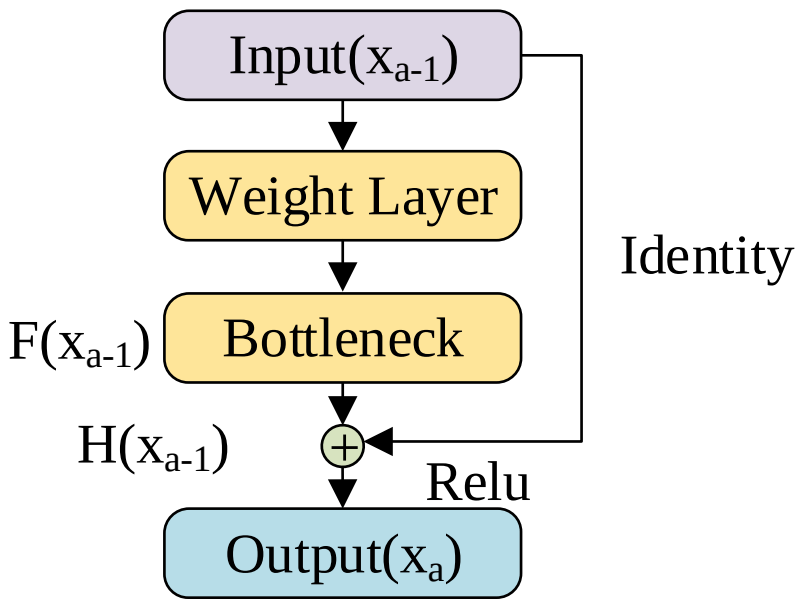

Fig. 5. Design principle of the ResNet.

In Fig. 5, $x_{a\text{-}1}$ is the input feature of the $a$-th layer of the network. It should be noted that ResNet has multiple variants, such as ResNet18, ResNet34, ResNet50, and ResNet101, with the number representing the number of layers in the model. $H(x_{a-1})$ represents the base mapping, expressed as:

$$H(x_{a-1}) = F(x_{a-1}) + x_{a-1} \tag{4}$$

where $H(x_{a-1})$ is the ideal mapping we hope the network to learn. However, ResNet does not directly learn this mapping but lets the network fit a residual mapping $F\left(x_{a-1}\right)$, expressed as:

$$F(x_{a-1}) = H(x_{a-1}) - x_{a-1} \tag{5}$$

From Equ. (5), it can be seen that learning the base mapping $H(x_{a-1})$ is equivalent to learning the residual mapping $F(x_{a-1})$. However, the residual mapping is usually easier to optimize, especially when the network is deep. This is because for identity mapping, the network only needs to learn the residual mapping as 0, while for nonlinear mapping, the network needs to fit a complex function.

The basic component unit of ResNet is the basic-block, and Fig.6 shows the structure composed of two basic-blocks.

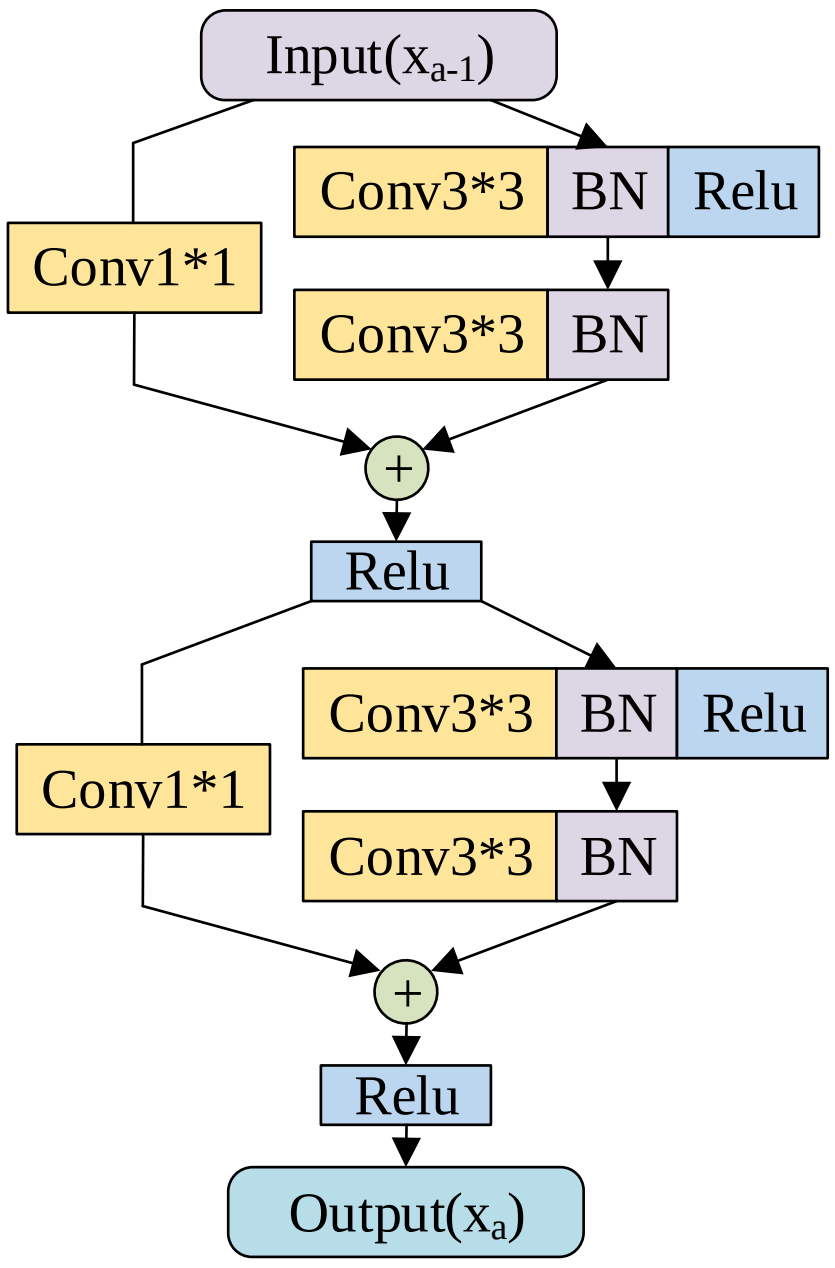

Fig. 6. Schematic diagram of two basic-blocks.

In Fig. 6, Conv represents convolution; BN is batch normalization; ReLU is the activation function. The mathematical expression of this residual idea is as:

$$H_a = f(F(x_{a-1}, \omega_{a-1}) + H'_{a-1}) \tag{6}$$

where $H_a$ represents the output feature of the residual structure; $F(x_{a-1}, \omega_{a-1})$ represents the residual mapping, $x_{a-1}$ represents the input feature of the $a$-th layer, $\omega_{a-1}$ represents the input weight of the $a$-th layer; $H'_{a-1}$ represents the nonlinear mapping of the previous residual feature to ensure that it can be added to the residual mapping; $f$ is the activation function.

Although ResNet can effectively solve the problem of gradient vanishing, for the task of batik image classification, the original ResNet still has some shortcomings. First, the residual blocks are weak in modeling capabilities for global and long-distance dependencies. Second, as the network deepens, the stacking of residual blocks may lead to the loss of resolution and semantic information of the feature maps. To address these issues, this paper proposes an improved ResNet model, as shown in Fig. 7. Considering the model's performance and computational efficiency, this paper selects ResNet34 as the basic architecture for improvement.

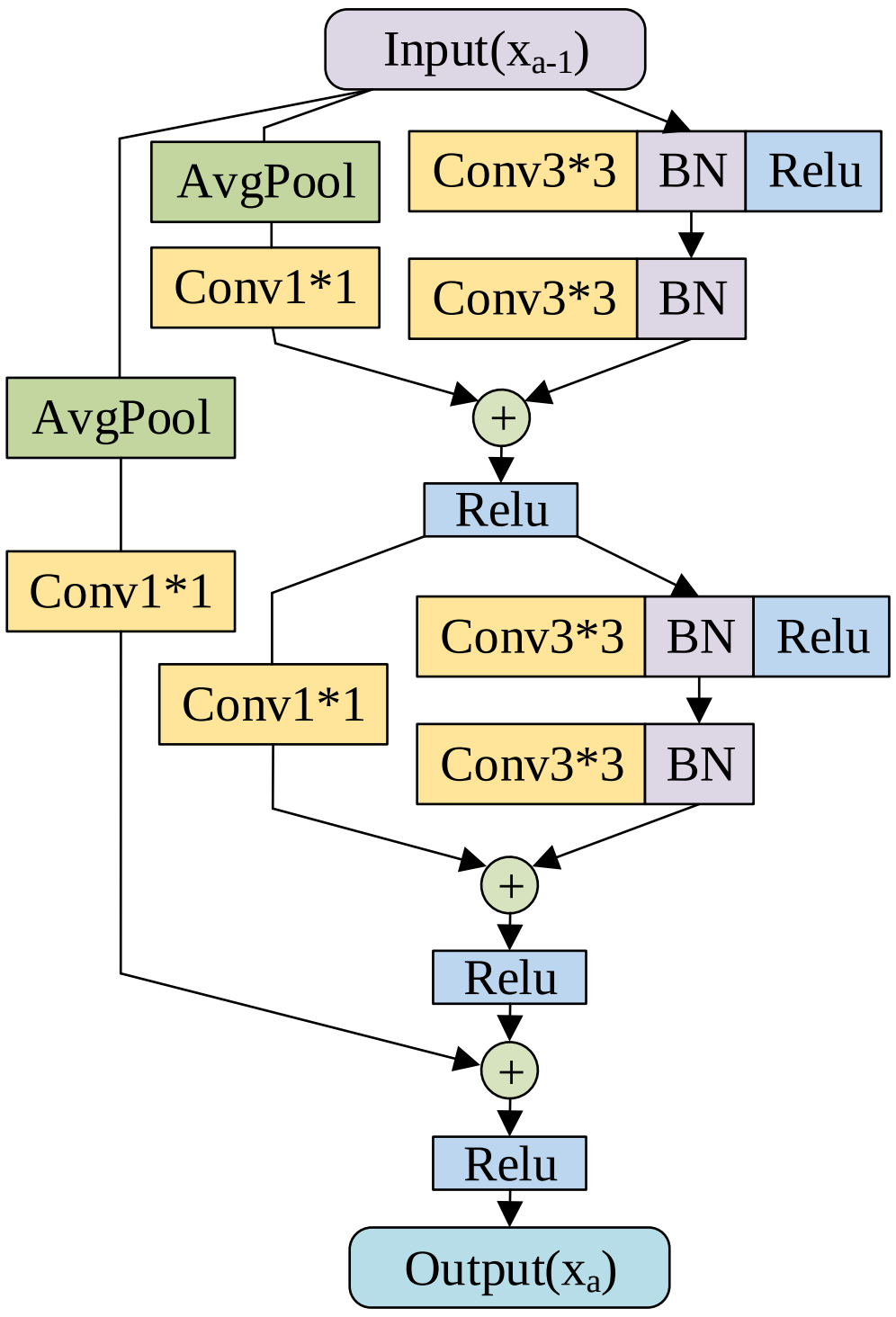

Fig. 7. Improved ResNet34 model.

Compared with the original ResNet34, the improvements in this paper are mainly reflected in two aspects:

(1) In the residual block, the 1*1 convolution is replaced by a combination of average pooling and convolution. The average pooling layer can reduce the resolution of the feature map, thus introducing multi-scale features. At the same time, the introduction of average pooling also increases the sparsity of the network, which helps to improve the model's generalization ability and robustness.

(2) After two residual blocks, a long-range residual connection is introduced to add the original input features to the output features of the residual block. This cross-layer feature reuse mechanism can help the network capture global and long-distance dependencies while alleviating the gradient vanishing problem.

Through the above improvements, the ResNet34 model in this paper can better adapt to the characteristics and requirements of the batik image classification task. On the one hand, the introduction of multi-scale features and the increase of sparsity enable the model to capture the rich texture and detail information in batik images. On the other hand, the addition of long-range residual connections enhances the model's ability to model global semantic information, which helps to distinguish similar batik categories. Fig. 8 shows the overall classification model framework used in this paper, where the input image size is 3*256*256.

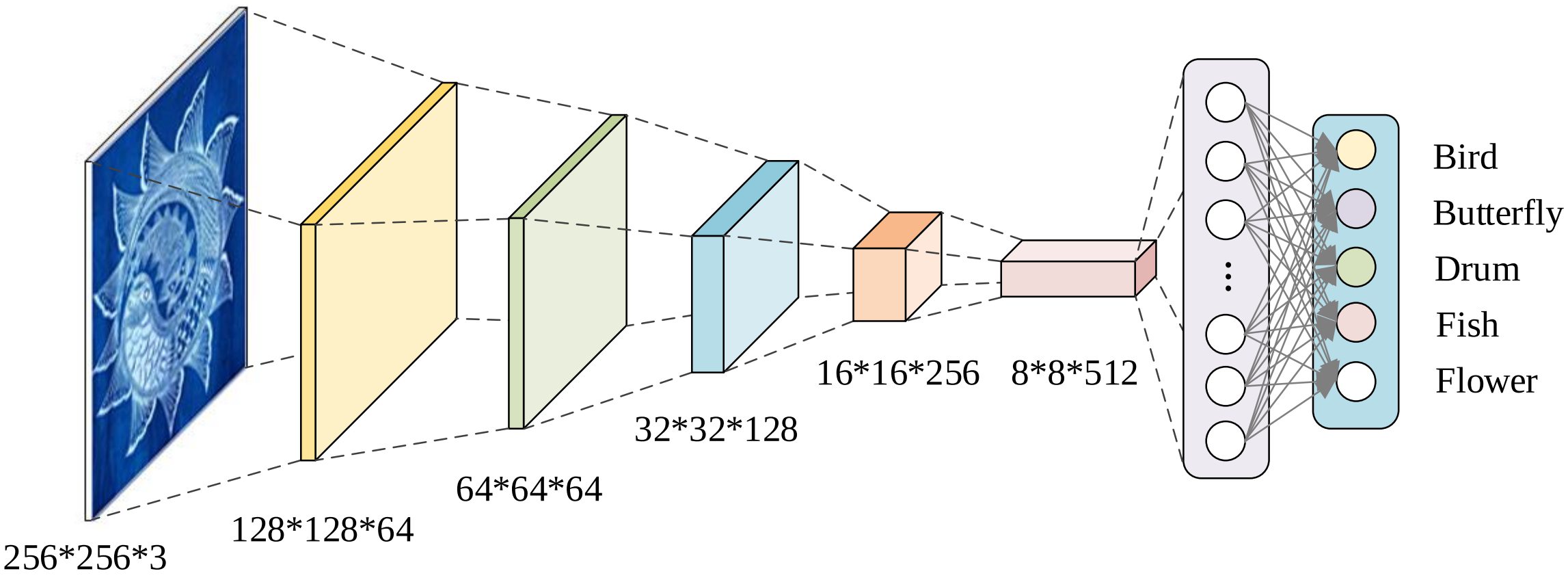

Fig. 8. Batik image classification model framework.

# 4. Experiments

## 4.1. BPKG Construction

### 4.1.1 Batik Ontology Model Construction

Ontology originates from philosophy and is later used in computer science to describe conceptual entities and their relationships. As a key component of the batik knowledge graph, the batik ontology can provide a clear structure and standardized semantics for the construction of the knowledge graph. We adopt the seven-step method [48] to construct the batik ontology model and define four ontology concepts: pattern, meaning, worship consciousness, and prototype source, as shown in Tab. 1. We also define nine relationships based on ontology concepts, such as mean, belong to, worship, origin from, etc. as shown in Tab. 2.

Tab. 1. Definition ontology concepts.

| Concept Name | Paraphrase | Concept Instance |
|---|---|---|
| 纹样(Pattern) | Describes the pattern in batik | 蝴蝶纹(butterfly pattern), 鱼纹(fish pattern) |
| 寓意(Meaning) | Describes the meaning of batik patterns | 子嗣绵延(praying for having children), 夫妻和睦(marital harmony) |
| 崇拜意识(Worship Consciousness) | Describes the worship aspect of batik patterns | 生殖崇拜(fertility worship), 鬼神崇拜(ghost-god worship) |
| 原型来源(Prototype Source) | Describes the prototype source of batik patterns | 神话故事(mythology), 自然化境(nature) |

Tab. 2. Definition of relationships between concepts.

| Relationship Name | Define domain | Value domain | Relationship instance |
|---|---|---|---|
| 蕴含<br>Mean | Pattern | Meaning | 蝴蝶纹-蕴含-敬重祖先<br>Butterfly -Mean-Respecting ancestor |
| 属于<br>Belong to | Pattern | Pattern | 鸟纹-属于-动物纹<br>Bird –Belong to-Animal |
| 崇拜<br>Worship | Pattern | Worship Consciousness | 鱼纹-崇拜-生殖崇拜<br>Fish -Worship-Fertility worship |
| 来源<br>Origin from | Pattern | Prototype Source | 梨花纹-来源于-大自然<br>Pear blossom- Origin from -Nature |
| 同义<br>Synonym | Pattern | Pattern | 蝴蝶纹-同义-蝴蝶妈妈<br>Butterfly -Synonym-Butterfly mom |
| 母子<br>Mother & child | Pattern | Pattern | 蝴蝶纹-母子-龙纹<br>Butterfly -Mother & child-Dragon |
| 父女<br>Father & daughter | Pattern | Pattern | 姜央-父女-相两<br>Jiang Yang-Father & daughter-Xiang Liang |
| 父子<br>Father & son | Pattern | Pattern | 姜央-父子-相芒<br>Jiang Yang-Father & son-Xiang Mang |
| 兄弟姐妹<br>Sibling | Pattern | Pattern | 相两-兄弟姐妹-相两<br>Xiang Mang-Sibling-Xiang Liang |

### 4.1.2 Batik Knowledge Extraction, Fusion, and Visualization

Considering the accuracy, richness, and scale of batik knowledge, we select literature and book data for research. We downloaded 100 journal papers and thesis from CNKI and obtained 30 e-books from the Chaoxing platform and the Internet. Subsequently, we extracted the text data from these documents and will use NLP techniques for analysis.

When processing Chinese text, word segmentation is required. By writing a Python program to call the Jieba library and adding a custom batik dictionary, the word segmentation effect is improved. For example, for the sentence "蝴蝶妈妈造就了苗族祖先姜央(The Butterfly mom created the Miao ancestors Jiang Yang)," if the dictionary is not added, the segmentation result is "蝴蝶/妈妈/造就/了/苗族/祖先/姜央" (Butterfly / Mom / created / the / Miao / ancestors / Jiang Yang), after adding the dictionary, the result becomes "蝴蝶妈妈/造就/了/苗族/祖先/姜央"(Butterfly Mother / created / the / Miao / ancestors / Jiang Yang). This more closely aligns with the actual context in Miao culture because "Butterfly Mom" is a complete term in their culture. Our batik dictionary contains 50 terms.

In the entity extraction task, we adopt a method based on the batik dictionary. Using the terms in the batik dictionary as center words, we perform clustering on the words in the text data based on cosine distance. Considering that Skip-gram has higher accuracy in word embedding training, this paper uses it for training. By calling the Word2vec model in the Gensim library and setting the corresponding parameters (as shown in Tab. 3), we obtain the embedding representations of the words.

Tab. 3. Word2vec parameter settings.

| Parameters | Setup |
|---|---|
| cbow | 0 |
| size | 200 |
| window | 5 |
| negative | 0 |
| hs | 1 |
| sample | 1e-4 |
| binary | 1 |
| alpha | 0.025 |
| min-count | 5 |
| batch-words | 10000 |

Based on the trained word embedding, we use the terms in the batik dictionary as center words and perform clustering on the words in the text data based on cosine distance, obtaining 50 clusters. Next, we manually filter and de-duplicate these 50 clusters, finally obtaining an entity set containing 120 terms. Compared with rule-based and machine learning-based methods, the dictionary-based method has the advantages of simple implementation and high computational efficiency. However, its disadvantage is the difficulty in discovering new entities outside the dictionary. To improve the coverage of entity extraction as much as possible, we further expand the entity set by manually filtering the clustering results on top of the initial clustering.

The relation extraction task is built upon entity extraction, aiming to extract the semantic relations between two or more entities from text data. In this research, we use the Language Technology Platform (LTP) developed by the Research Center for Social Computing and Information Retrieval at Harbin Institute of Technology for relation extraction. LTP provides various Chinese NLP functions, among which the dependency parsing module can automatically identify the dependency relations between words in a sentence, providing an important grammatical basis for relation extraction.

Specifically, we first input the entity pairs (entity1, entity2) obtained from entity extraction and their corresponding sentences into LTP. For example, for the sentence "石榴纹是一种常见植物纹" (Pomegranate pattern is a common plant pattern) containing the entities "pomegranate pattern" and "plant pattern", we input it into LTP's dependency parsing module. The module automatically splits the sentence into the words "石榴纹"(pomegranate pattern), "是"(is), "一种"(a), "常见"(common), and "植物纹"(plant pattern), and performs part-of-speech tagging for each word. Next, LTP identifies the dependency relations between the words in the sentence based on predefined Chinese dependency parsing rules. For example, the rule "if a noun forms a subject-predicate relationship with '是'(is), then the noun is the subject" can be used to identify the subject-predicate relationship between "pomegranate pattern" and "is". Similarly, the rule "if there exists an 'is' predicate between two nouns, and the second noun has a verb-object relationship with 'is', then the two nouns have an 'is' relationship" can identify the relationship between "pomegranate pattern" and "plant pattern".

Through the above analysis, LTP obtains the dependency parsing tree of the sentence, as shown in Fig. 7. It can be seen from the figure that there exists an SBV (subject-predicate relationship) between "pomegranate pattern" and "is", and a VOB (verb-object relationship) between "is" and "plant pattern". According to our predefined relation extraction rule, i.e., "if two entities in a sentence are connected through SBV and VOB relationships, then there exists an 'is' relationship between these two entities", we extract the triple (pomegranate pattern, is, plant pattern) from this parsing tree.

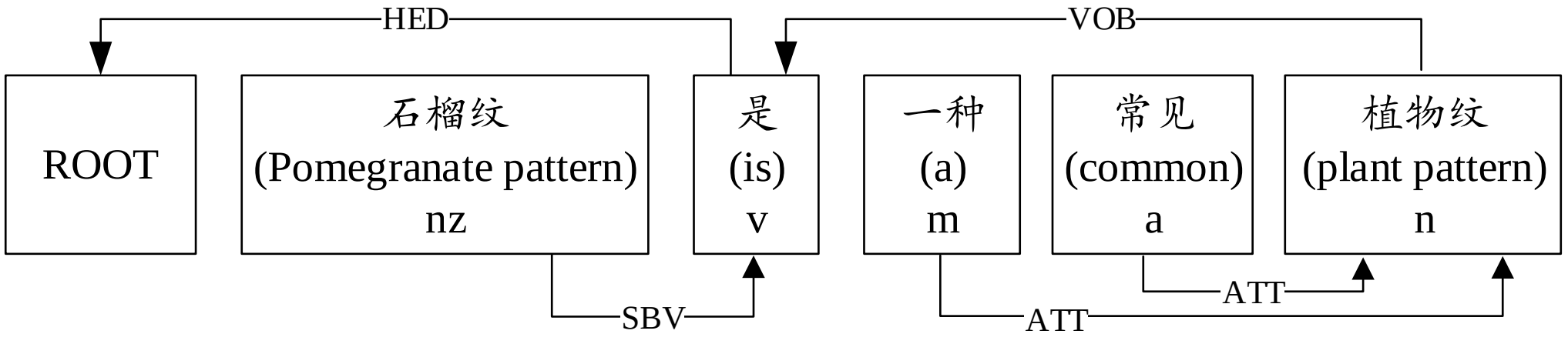

Fig.7 Dependency parsing results.

Furthermore, we perform normalization processing on the extracted triples based on the entity attributes and relation patterns defined in the ontology. For example, the ontology stipulates that "plant pattern" is a pattern category, and if a pattern entity has an "is" relationship with it, then it can be converted to a "belong to" relationship. Therefore, we finally obtain the normalized triple (pomegranate pattern, belong to, plant pattern).

We repeat the above steps for all sentences containing entity pairs, obtaining a large number of triples. However, because some sentences have similar grammatical structures, the triples they generate may be duplicate; moreover, for the same entity pair, the same or similar semantic relations may be expressed in different sentences, resulting in the existence of multi-valued triples. To solve this problem, we adopt a de-duplication strategy, deleting completely duplicate triples; for multi-valued triples, we perform de-duplication through manual filtering, i.e., judging whether different triples express the same semantics, and keeping only one if they do.

After a series of processing steps, we finally obtain 419 triples. To intuitively display these complex semantic relations, we use the Neo4j graph database for storage and generate the BPKG shown in Fig. 8. In this knowledge graph, nodes represent batik pattern entities, edges represent relations between entities, and different types of relations are distinguished by edges of different colors. Through this graphical method, we can more clearly and intuitively display the connections between batik patterns, which helps to better understand and apply this knowledge.

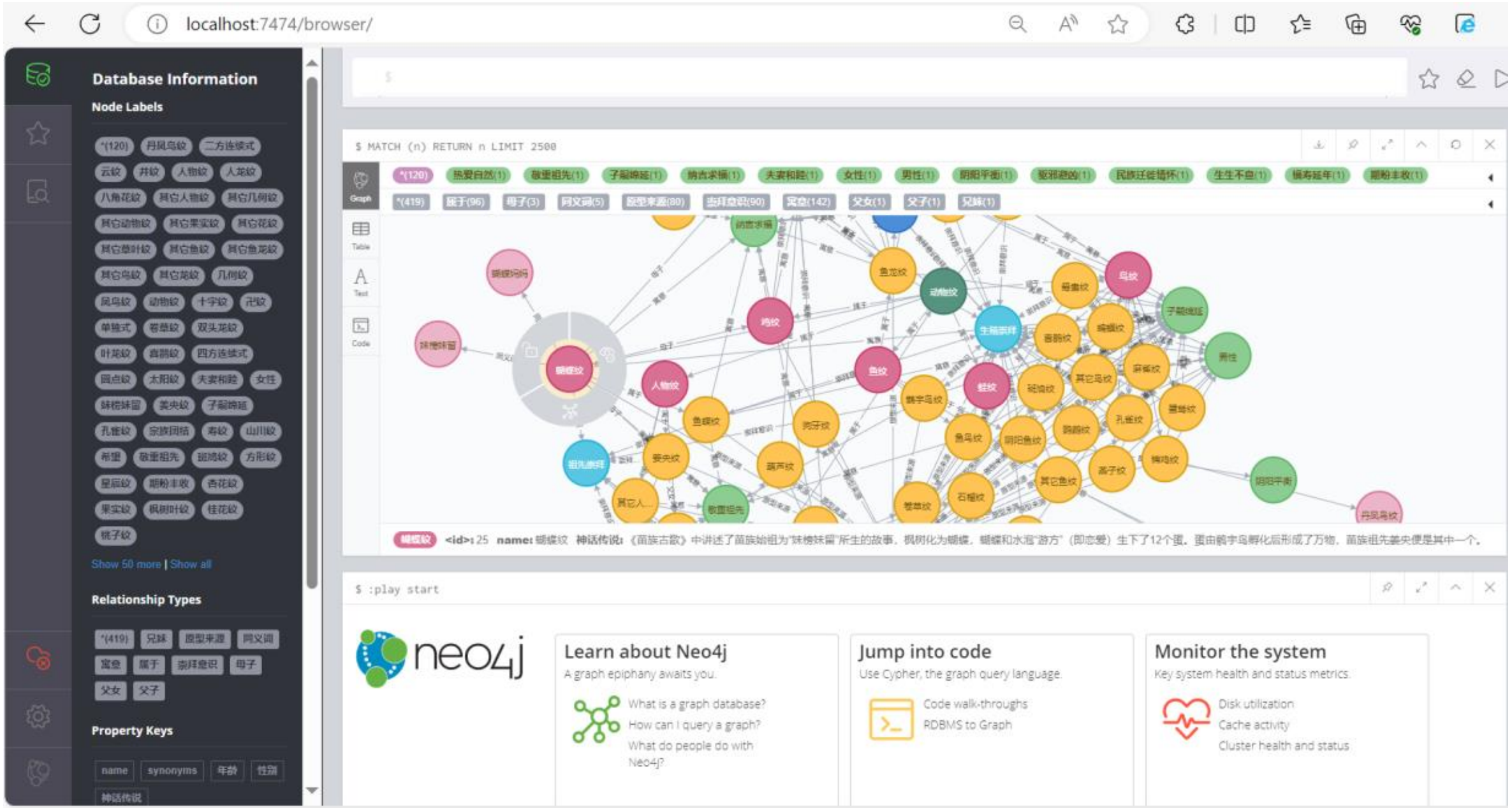

Fig. 8. The structure of BPKG . Nodes of different colors represent different types of entities, and the directed edges between nodes represent the relationships between entities.

## 4.1.3 Semantic Retrieval

Semantic retrieval of the BPKG can be realized through Cypher, a declarative query language for querying and manipulating graph databases. Users can express complex semantic query needs by defining the matching of nodes and relations. In Cypher, nodes are represented by parentheses "( )", and relations are represented by square brackets "[ ]". For example, ($n_i$)-[$r$]->($n_j$) represents that node $n_i$ is connected to node $n_j$ by relation $r$. The commonly used clauses in Cypher include *Match*, *Where*, and *Return*. Among them, *Match* defines the basic matching pattern of the query, and is applicable to both nodes and relations. *Where* specifies the matching conditions, usually limiting the properties of the variables appearing in the match. *Return* specifies the returned results.

For example, if users want to query all pattern types in the BPKG, they can use the following code: $ MATCH (n)-[r:属于]->(m) RETURN n, r, m. This query means to match all node pairs ($n$, $m$) that exist in the "属于"(belong to) relationship and return them. $n$ and $m$ represent the start and end points of the "belong to" relation, respectively, and $r$ represents the relation itself. The query results are shown in Fig. 9. Through this hierarchical classification system, users can quickly understand the classification system of batik patterns and browse the specific patterns under each category.

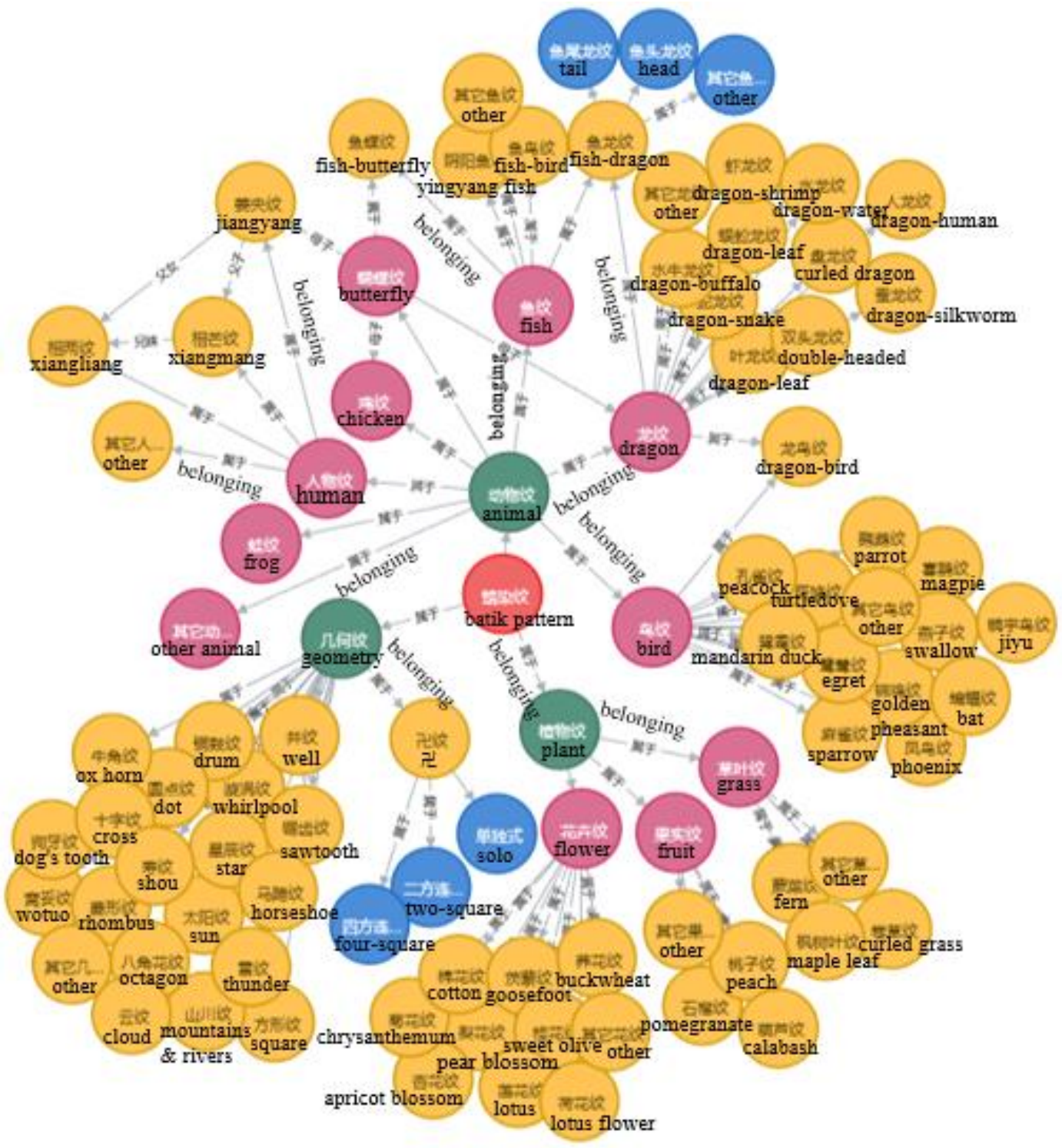

Fig. 9. Belong to relationship. In BPKG, we divide batik patterns into three main categories: animal, geometric, and plant patterns. Among them, animal patterns are further divided into eight subcategories: butterfly, fish, bird, dragon, frog, chicken, human, and other animal patterns, containing a total of 43 specific patterns. Plant patterns are divided into three subcategories: flower, fruit, and leaf patterns, containing a total of 18 patterns. Geometric patterns are divided into 23 subcategories such as drum, vortex, and dot patterns.

In addition to querying the classification of patterns, users can also retrieve based on the meaning of patterns. For example, if one wants to query all patterns that imply praying for having children, the following code can be executed: $ MATCH p=()-[r:寓意]->(m:子嗣绵延) RETURN p. This query will match all pattern nodes that have an "子嗣绵延" (praying for having children) relationship and return the results. Similarly, if a user wants to query the meaning of all patterns, the query statement is: $ MATCH (n)-[r:寓意]->(m) RETURN n, r, m. The query results are shown in Fig. 10. It can be seen that each batik pattern contains specific cultural meanings. For example, the butterfly represents "respecting ancestor", the pear blossom mean "hope", and the pomegranate pattern has both "loving nature" and " praying for having children". Through BPKG, these cultural connotations hidden behind the patterns are visualized, helping users to better understand and inherit batik culture.

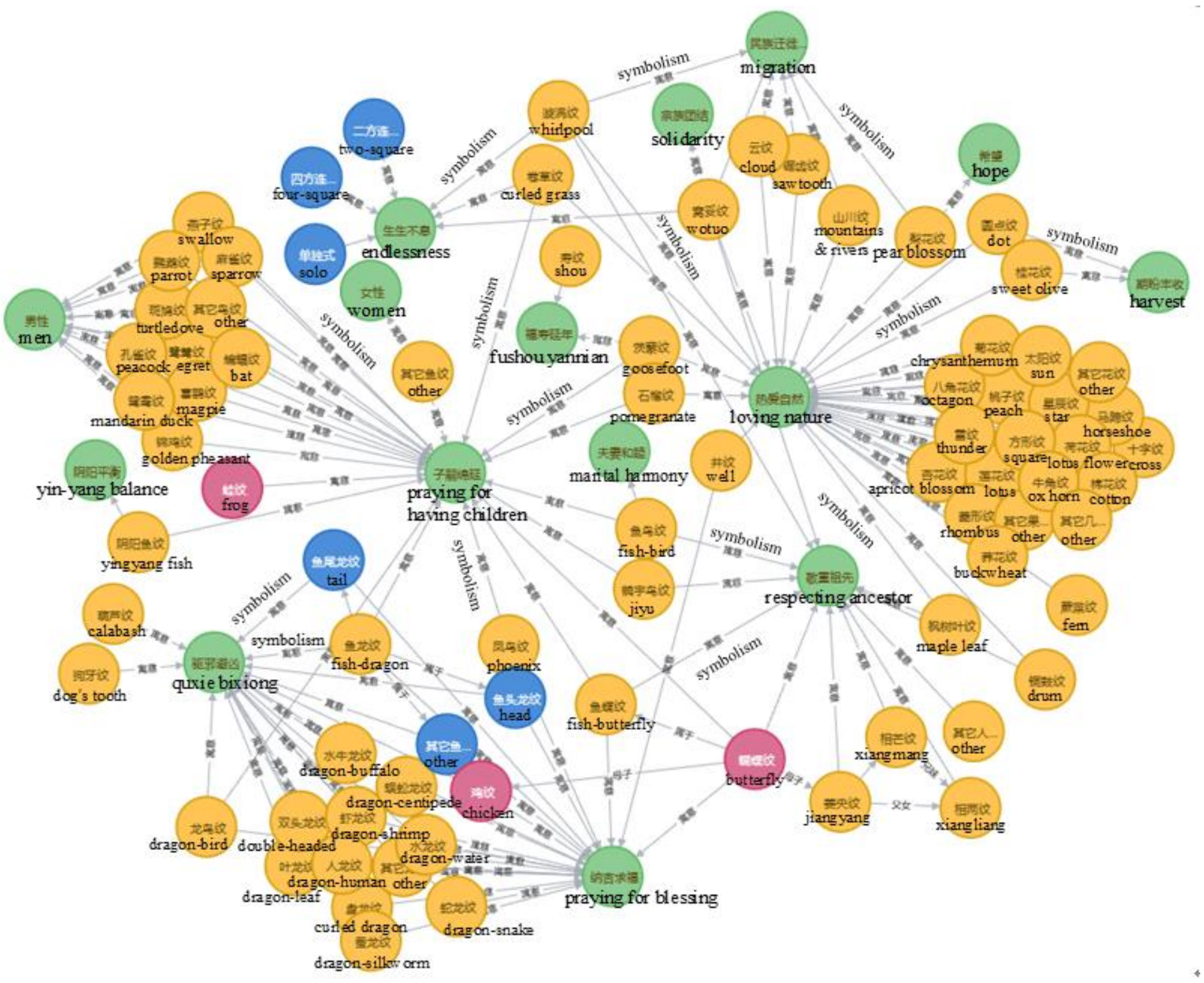

Fig. 10. Mean relationship. We summarize the meanings of batik patterns into 15 categories: 热爱自然(loving nature), 敬重祖先(respecting ancestor), 子嗣绵延(praying for having children), 纳吉求福(praying for blessing), 夫妻和睦(marital harmony), 代表女性(women), 代表男性(men), 阴阳平衡(yin-yang balance), 趋吉避凶(quxie bixiong), 民族迁徙情怀(migration), 生生不息(endlessness), 福寿延年(fushou yannian), 期盼丰收(harvest), 宗族团结(solidarity), and 希望(hope).

In summary, we construct a BPKG containing 120 entities and 419 triples, covering multiple dimensions of information such as the classification, meaning, and worship, with high knowledge coverage and semantic richness. Through the Cypher query language, users can perform semantic retrieval to obtain the patterns and their semantic information of interest. This knowledge not only helps the digital protection of batik culture but also provides more materials and inspiration for the design and application of batik patterns.

## 4.2. Pattern Classification

### 4.2.1 Dataset

To construct the batik pattern dataset, we visited the Qiandongnan Miao and Dong Autonomous Prefecture in Guizhou Province, China, and collected more than 1,000 original images of batik objects using professional equipment. In the image preprocessing stage, we first manually filtered the original images based on the completeness and clarity of the patterns, removing some low-quality samples. Then, we used Photoshop software to extract independent pattern samples from the original images, with the pattern boundaries as the cropping regions. Finally, we obtained a batik pattern dataset containing 12,249 images. This dataset covers the five most representative patterns in batik, namely bird(2,853 images), butterfly(2,064 images), drum (2,065 images), fish (2,122 images), and plant(3,145 images) patterns. Some sample images are shown in Fig. 11.

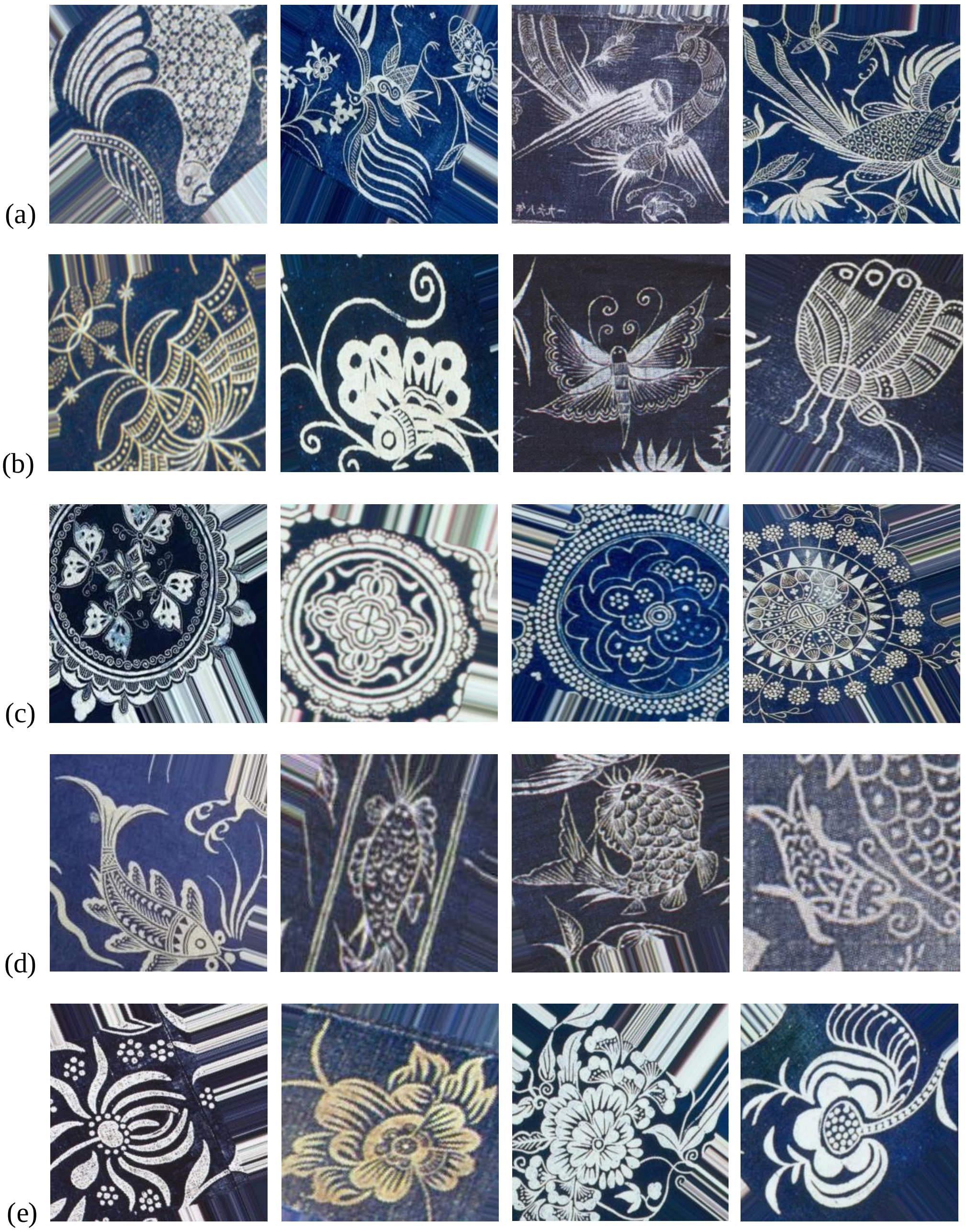

Fig. 11. Sample examples of the batik pattern dataset. (a) bird; (b) butterfly; (c) fish; (d) drum; (e) plant.

Due to differences in acquisition conditions and cropping regions, the samples in the dataset show significant heterogeneity in size, color, texture, and other aspects. At the same time, some samples have large inter-class differences and small intra-class differences, which will interfere with the model's feature extraction and classification. To reduce interference, we normalize all images to a size of 3*256*256, which can largely retain the detailed features of the patterns without placing too high hardware requirements on the deep learning model.

### 4.2.2 Experimental Environment

The hardware and software configurations in the experiments are as follows: Intel(R) Core(TM) i7-11800H processor, 8 cores and 16 threads, main frequency 2.3GHz, 16GB running memory, graphics processor GeForce RTX 3080, 12GB video memory; deep learning model framework uses Pytorch1.9.1 and Torchvision 0.10.1.

To evaluate the performance of the proposed model, we select three classic CNN as benchmark models, namely AlexNet, VGG16, and ResNet34. Among them, AlexNet is the first deep network structure that achieved breakthrough results in large-scale image classification tasks, promoting the development of deep learning in computer vision; VGG16 increases the depth of the network while keeping the receptive field unchanged by using a series of small-sized convolution kernels, thereby improving the ability of feature representation; ResNet34 effectively solves the problem of difficult training of deep networks by introducing residual connections, making the training of ultra-deep networks possible. These three models have been widely used and validated in many visual tasks such as image classification and object detection, making them very suitable as benchmarks for measuring the performance of new models. In the experiments, to ensure fairness of comparison, we do not use the pre-trained weights of these models on other large-scale datasets, but train them from scratch on our own dataset, making their starting state consistent with our model.

We randomly divide the dataset into training and test sets in a ratio of 6:4, i.e., 7,348 training samples and 4,901 test samples. This division ratio is a common empirical value, which can ensure a sufficient number of training samples while leaving enough samples for performance evaluation in the test set. The training parameters of our model are shown in Tab. 4, using the Adam optimizer for training, with 100 iterations and a batch size of 8.

Tab. 4 Training parameter setting

| Parameters | Setup |
|---|---|
| Epochs | 100 |
| Batch Size | 8 |
| Optimizer | Adam |
| Initial Learning Rate | 1e-3 |
| Final Learning Rate | 1e-4 |
| Momentum | 0.9 |
| Loss | CrossEntropy |
| RandomResizedCrop | Open |
| RandomHorizontalFlip | Open |

4.2.3 Experimental Results and Analysis

To comprehensively evaluate the performance of the classification model, we adopt four commonly used indicators: accuracy, precision, recall, and F1-score. The calculation formulas of these indicators are as follows:

$$Accuracy = \frac{TP + TN}{TP + TN + FP + FN} \tag{7}$$

$$Precision = \frac{TP}{TP + FP} \tag{8}$$

$$Recall = \frac{TP}{TP + FN} \tag{9}$$

$$F1 - Score = \frac{2 \times Recall \times Precision}{Recall + Precision} \tag{10}$$

where $TP$ represents the number of samples that the label is true and the prediction result is also true; $TN$ represents that the label is false and the prediction result is also false; $FP$ represents that the label is false and the prediction result is true; and $FN$ represents that the label is true and the prediction result is false.

Inputting the test set samples into the four trained classification models, we obtain the confusion matrices shown in Fig. 12. It can be seen that different models show different advantages on the task of classifying five batik patterns.

From an overall perspective, there are some differences in the classification effects of the five patterns. We calculate the accuracy of different patterns based on the confusion matrices, as shown in Tab. 5. Among them, the drum is the best, with over 98% accuracy on all four models. The reason for this phenomenon may be that the texture features of drum are more prominent, and the differences between samples are smaller, making them easier to recognize. In comparison, the classification effects of butterfly and fish are slightly lower, with accuracy below 86% on AlexNet and VGG16. This may be because the shape and texture features of these two types of patterns are more diverse, and the differences between samples are larger, posing greater challenges to feature extraction and classification. In addition, butterfly contain some detailed features, such as antennae and wing textures, which may be ignored by shallow networks, affecting the classification effect.

Specifically, in the bird classification, ResNet34 performs best, with 1,134 correctly classified samples, and our model closely follows, correctly classifying 1,133 samples, only one less than ResNet34. In contrast, AlexNet performs the worst, correctly classifying only 985 samples. In the classification of butterfly, our model achieves the highest number of correctly classified samples (813), outperforming the other three models. AlexNet performs the worst, with only 686 correctly classified samples. In drum classification, our model and ResNet34 both achieve the best performance, with the number of correctly classified samples reaching 822. In comparison, AlexNet performs the worst, with 814 correctly classified samples. In fish classification, ResNet34 ranks first with 838 correctly classified samples, and our model closely follows with 836 correctly classified samples, with a very small gap between the two. VGG16 performs the worst on this task, with 729 correctly classified samples. In plant classification, our model far outperforms the other three models with 1,247 correctly classified samples, showing a clear advantage. VGG16 performs the worst on this task, with 1,155 correctly classified samples. The above results indicate that our model and ResNet34 both show the best performance on the classification tasks of the five patterns, followed by the VGG16, while the AlexNet performs the worst.

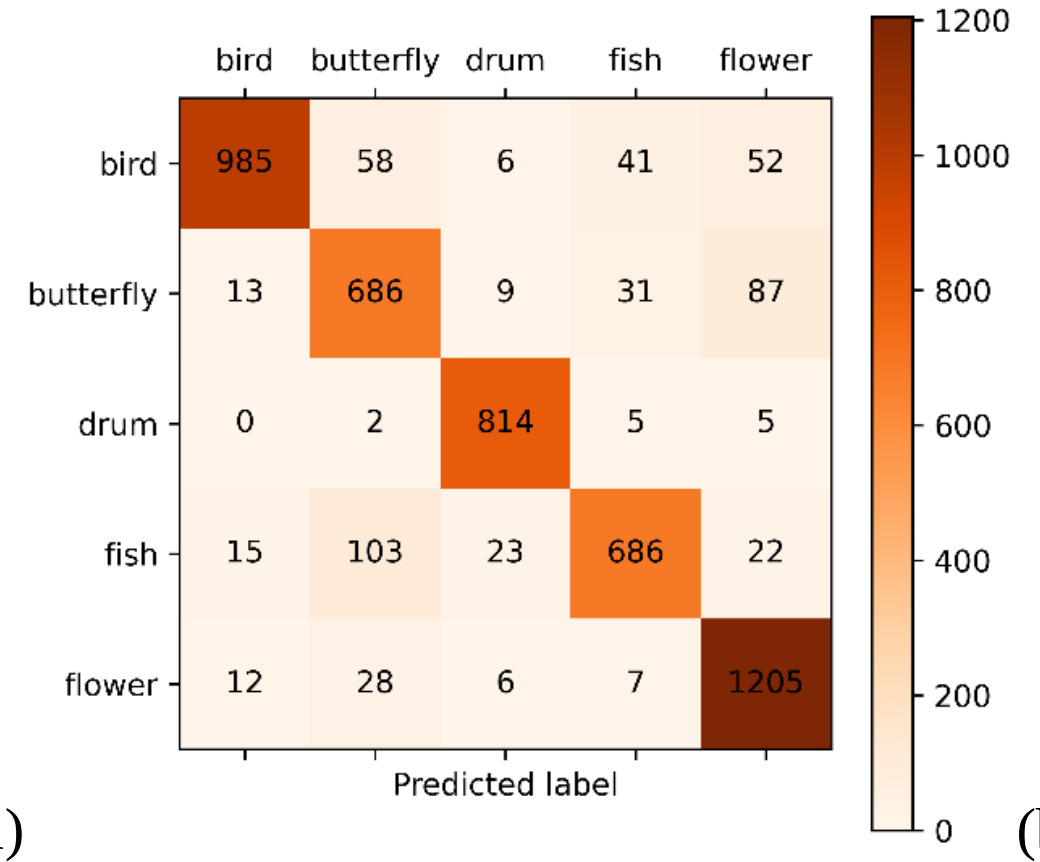

(a)

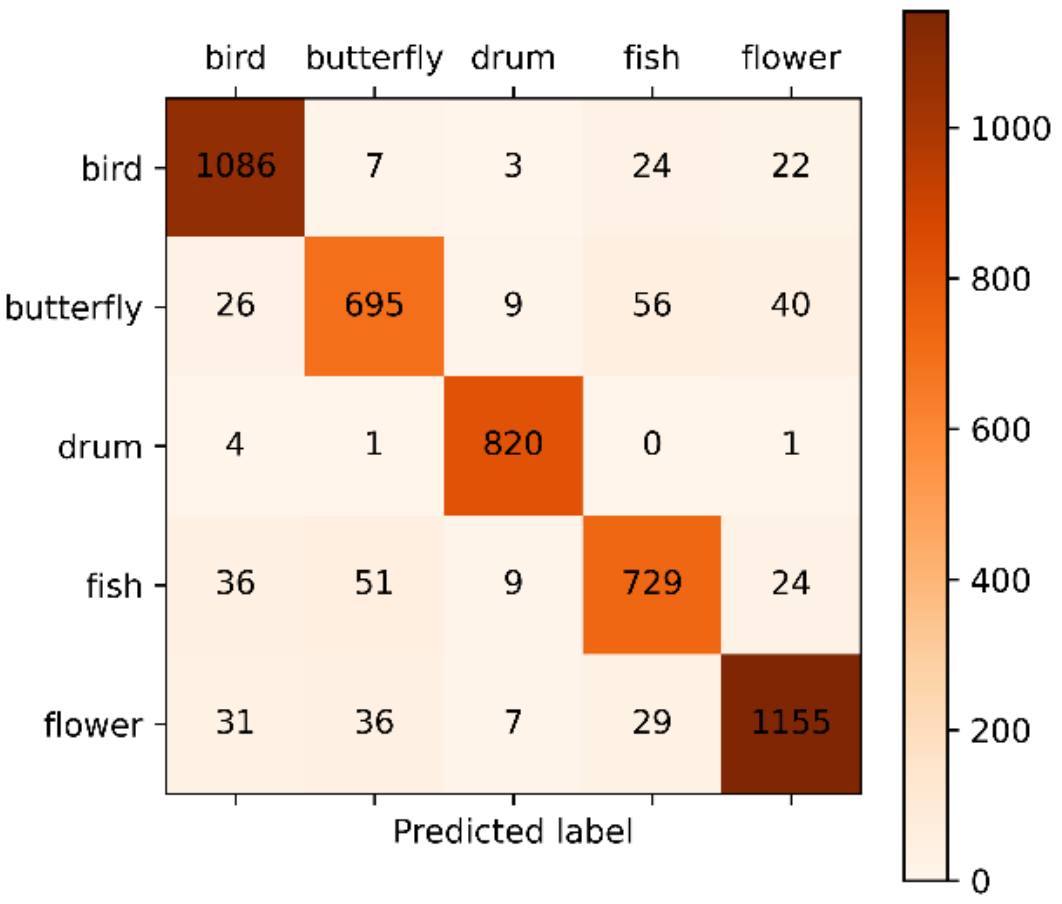

(b)

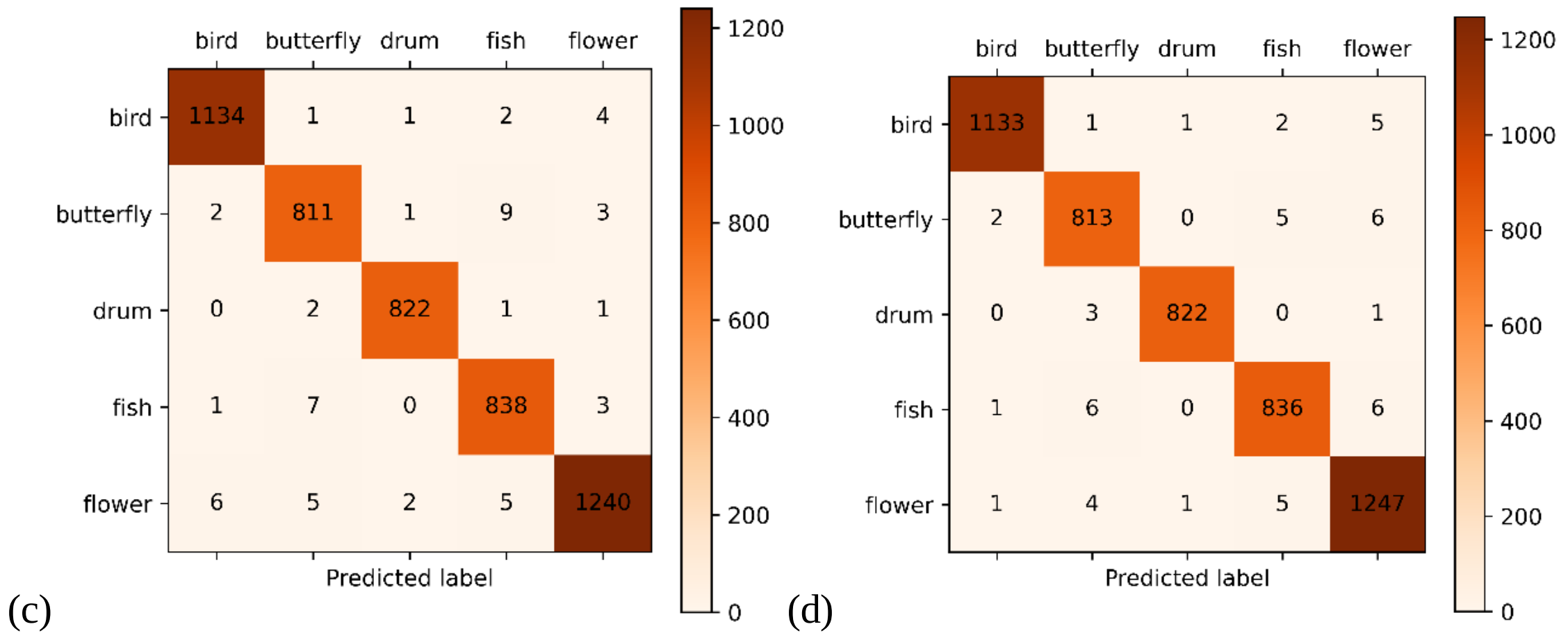

Fig. 12. Confusion matrix diagrams of classification models. (a) AlexNet; (b) VGG16; (c) ResNet34; (d) our model.

Tab. 5. Accuracy comparison of different patterns.

| | **AlexNet** | **VGG16** | **ResNet34** | **Ours** |
|---|---|---|---|---|
| **bird** | 0.8625 | 0.9510 | 0.9930 | 0.9921 |
| **butterfly** | 0.8305 | 0.8414 | 0.9818 | 0.9843 |
| **drum** | 0.9855 | 0.9927 | 0.9952 | 0.9952 |
| **fish** | 0.8080 | 0.8587 | 0.9870 | 0.9847 |
| **flower** | 0.9579 | 0.9181 | 0.9857 | 0.9913 |

To evaluate the performance of each model under different classification thresholds, we plot the ROC curves shown in Fig. 13. The ROC curve reflects the relationship between the true positive rate (TPR) and the false positive rate (FPR) under different classification thresholds. Among them, TPR represents the proportion of correctly classified positive samples among all positive samples, and FPR represents the proportion of incorrectly classified negative samples among all negative samples. The closer the curve is to the upper left corner, the better the classification performance of the model. The area under the curve (AUC) is a commonly used comprehensive evaluation indicator, with values ranging from 0 to 1, and larger values indicating better classifier performance.

From Fig. 13, we can see that our model achieves excellent performance on all patterns, with its ROC curve being the closest to the upper left corner and having the highest AUC value of 0.99. ResNet34 also performed well, with an average AUC value of 0.98. In comparison, the ROC curves of VGG16 and AlexNet are relatively farther from the upper left corner, with average AUC values of 0.96 and 0.95, respectively, indicating their relatively poorer classification performance. Especially in the butterfly classification, the advantages of our model and ResNet34 are more obvious. This may be because butterfly contain a large number of detailed texture features, and our model, by fusing local and global information, better captures and utilizes these key features, thus obtaining more accurate classification results.

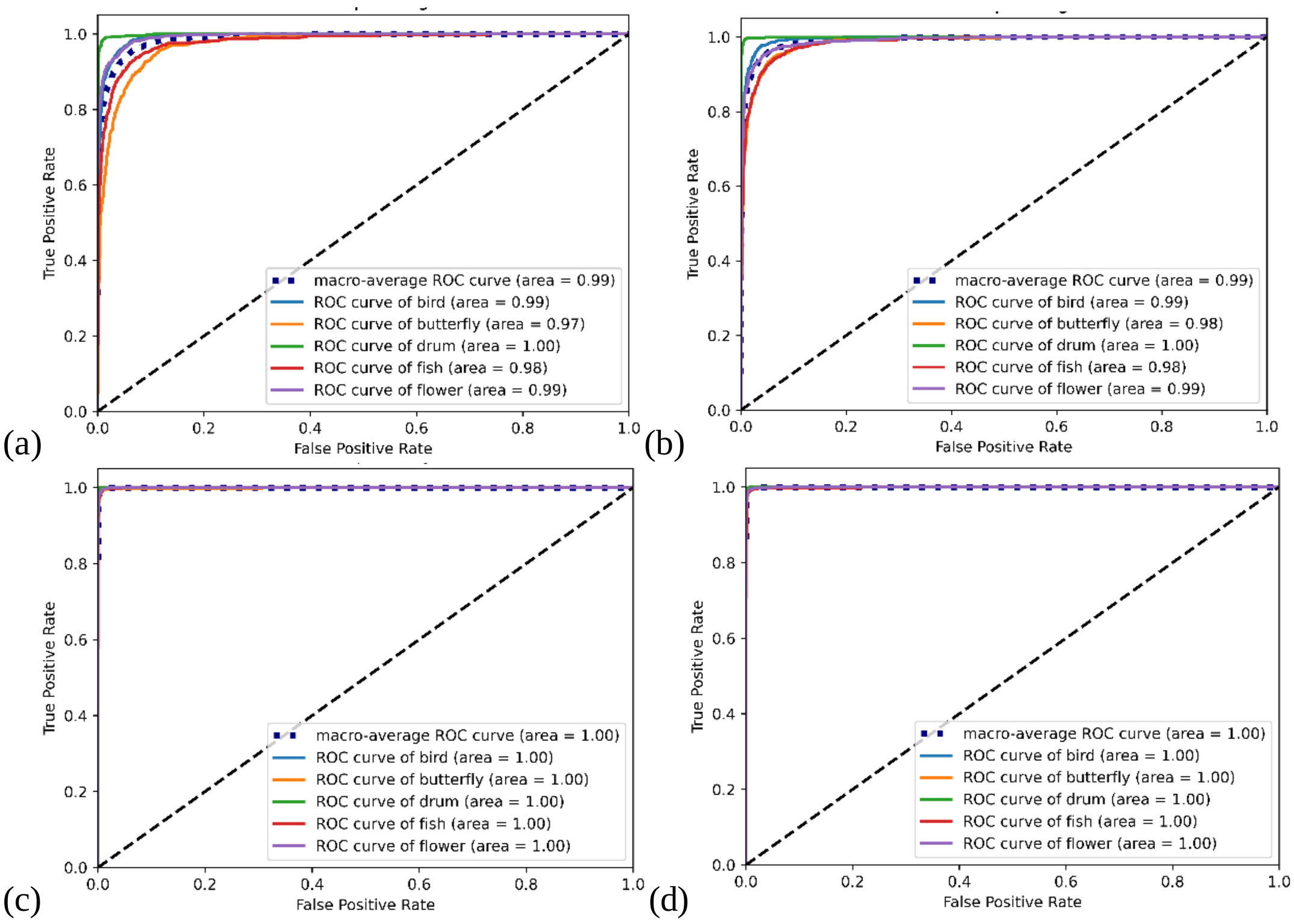

Fig. 13. ROC curves of classification models. (a) AlexNet; (b) VGG16; (c) ResNet34; (d) Our model.

To further evaluate the performance of the four models, we calculate their accuracy, precision, recall, and F1-score, with detailed data shown in Tab. 6. Our model achieves the best results on all indicators, with accuracy, precision, and F1-score reaching 0.99, and recall reaching 0.989. ResNet34's performance ranked second, with values above 0.986 on all four indicators. The performance of the VGG16 is lower than ResNet34 but better than AlexNet. AlexNet performs the worst, with values below 0.90 on all four indicators. These results further verify the effectiveness and superiority of our model on the batik pattern classification task.

Tab. 6. Comparison of classification indicators.

| Indicators | AlexNet | VGG16 | ResNet34 | Ours |
|---|---|---|---|---|
| Accuracy | **0.893** | 0.915 | 0.986 | **0.990** |
| Precision | **0.892** | 0.913 | 0.988 | **0.990** |
| Recall | **0.889** | 0.912 | 0.988 | **0.989** |
| F1-Score | **0.889** | 0.912 | 0.988 | **0.990** |

## 5. Conclusion

Batik, as a unique decorative art, has played an important role in the history of Chinese ethnic minorities. Batik patterns not only have aesthetic value but also carry the cultural memory and emotional identity of the Miao people. However, in the changes of modern society, batik culture is facing the dilemma of insufficient inheritance and accelerated loss.

To promote the dissemination and protection of Chinese batik, this paper comprehensively applies knowledge graph, NLP, deep learning, and other artificial intelligence technologies to construct a dual-channel mechanism connecting batik images and cultural knowledge. First, through NLP techniques, entities and relationships are extracted from a large number of documents, and stored and visualized with the Neo4j database. Second, an improved ResNet34 model is proposed to improve the accuracy of image classification. Finally, the improved model is applied to the automatic classification of batik images, providing a tool for understanding batik images.

This research reduced the difficulty for non-professional users to understand and apply batik knowledge, opening up new paths for the cross-border dissemination and innovative application of batik culture. Whether they are professional designers or the general public interested in batik, they can use this mechanism to cognize, understand, appreciate, and innovate batik from different perspectives and at different levels. In future research, we will strive to further expand the scale of BPKG, enrich its semantic associations, optimize the retrieval and question-answering capabilities, and develop a cultural popularization platform, hoping to provide more reliable theoretical foundations and practical support for the dissemination and development of cultural heritage.

**Author Contributions:** H.Q. and Y.L. designed the study, performed the experiments, analyzed the results, and wrote the main manuscript. D.L. provided valuable insights and suggestions on the methodology, and knowledge graph construction. Y.Z. prepared the dataset used in the study. All authors reviewed the manuscript.

**Acknowledgments:** This work is supported by Guizhou Provincial Basic Research Program (Natural Science) under grant No. ZK[2023]029.

**Conflicts of Interest:** The authors declare no conflict of interest.